\let\useblackboard=\iftrue
%
%
\newfam\black
\input harvmac.tex
\noblackbox
\def\Title#1#2{\rightline{#1}
\ifx\answ\bigans\nopagenumbers\pageno0\vskip1in%
\baselineskip 15pt plus 1pt minus 1pt
\else
\def\listrefs{\footatend\vskip 1in\immediate\closeout\rfile\writestoppt
\baselineskip=14pt\centerline{{\bf References}}\bigskip{\frenchspacing%
\parindent=20pt\escapechar=` \input
refs.tmp\vfill\eject}\nonfrenchspacing}
\pageno1\vskip.8in\fi \centerline{\titlefont #2}\vskip .5in}

\ifx\answ\bigans\def\tcbreak#1{}\else\def\tcbreak#1{\cr&{#1}}\fi
\useblackboard
\message{If you do not have msbm (blackboard bold) fonts,}
\message{change the option at the top of the tex file.}

\font\blackboard=msbm10 scaled \magstep1
\font\blackboards=msbm7
\font\blackboardss=msbm5
\textfont\black=\blackboard
\scriptfont\black=\blackboards
\scriptscriptfont\black=\blackboardss

\else

\fi
%
\def\yboxit#1#2{\vbox{\hrule height #1 \hbox{\vrule width #1
\vbox{#2}\vrule width #1 }\hrule height #1 }}
\def\fillbox#1{\hbox to #1{\vbox to #1{\vfil}\hfil}}
\def\ybox{{\lower 1.3pt \yboxit{0.4pt}{\fillbox{8pt}}\hskip-0.2pt}}

\def\comments#1{}

\def\half{{1\over 2}}

\def\II{\relax{I\kern-.07em I}}

\def\inbar{\,\vrule height1.5ex width.4pt depth0pt}
\def\IZ{\relax\ifmmode\mathchoice
{\hbox{\cmss Z\kern-.4em Z}}{\hbox{\cmss Z\kern-.4em Z}}
{\lower.9pt\hbox{\cmsss Z\kern-.4em Z}}
{\lower1.2pt\hbox{\cmsss Z\kern-.4em Z}}\else{\cmss Z\kern-.4em
Z}\fi}
\def\IB{\relax{\rm I\kern-.18em B}}
\def\IC{{\relax\hbox{$\inbar\kern-.3em{\rm C}$}}}
\def\ID{\relax{\rm I\kern-.18em D}}
\def\IE{\relax{\rm I\kern-.18em E}}
\def\IF{\relax{\rm I\kern-.18em F}}
\def\IG{\relax\hbox{$\inbar\kern-.3em{\rm G}$}}
\def\IGa{\relax\hbox{${\rm I}\kern-.18em\Gamma$}}
\def\IH{\relax{\rm I\kern-.18em H}}
\def\II{\relax{\rm I\kern-.18em I}}
\def\IK{\relax{\rm I\kern-.18em K}}
\def\IP{\relax{\rm I\kern-.18em P}}

\font\cmss=cmss10 \font\cmsss=cmss10 at 7pt
\def\IR{\relax{\rm I\kern-.18em R}}

\def\BZ{\IZ}

%
%

\def\NP{{\it Nucl. Phys.\ }}
\def\AP{{\it Ann. Phys.\ }}

\def\PR{{\it Phys. Rev.\ }}
\def\PRL{{\it Phys. Rev. Lett.\ }}
\def\CMP{{\it Comm. Math. Phys.\ }}
\def\JMP{{\it J. Math. Phys.\ }}

\def\IJMP{{\it Int. Jour. Mod. Phys.\ }}
\def\Mod{{\it Mod. Phys. Lett.\ }}

\Title{ \vbox{\baselineskip12pt\hbox{hep-th/9702006}
\hbox{CALT-68-2096}
}}
{\vbox{
\centerline{Bound States of Type I$'$ D-particles}
\centerline{and Enhanced Gauge Symmetry}}}

\centerline{David A. Lowe}
\medskip
\centerline{California Institute of Technology}
\centerline{Pasadena, CA  91125, USA}
\centerline{\tt lowe@theory.caltech.edu}
\bigskip
Duality between the $E_8\times E_8$ heterotic string and Type I$'$
theory predicts a tower of D(irichlet)-particle bound states corresponding
to perturbative heterotic string states.
In the limit of infinite Type I$'$ coupling, some of
these bound states become massless, giving rise to enhanced $E_8
\times E_8$ gauge symmetry. By taking a different infinite coupling
limit,
one can recover the $E_8\times E_8$ gauge bosons of M-theory,
compactified on $S^1/\BZ_2$. In this paper we use the matrix model
description of the D-particle dynamics to study these bound states.
We find results consistent with the chain of dualities and
clarify a number of issues that arise in the application of the
matrix mechanics to this system.

\Date{January 1997}

\lref\daniel{U. Danielsson and G. Ferretti, hep-th/9610082.}
\lref\polwit{J. Polchinski and E. Witten, \NP {\bf B460} (1996) 525,
hep-th/9510169.}
\lref\kacsil{S. Kachru and E. Silverstein, hep-th/9612162.}
\lref\ginsparg{P. Ginsparg, \PR {\bf D35} (1987) 648.}
\lref\danieltwo{U. Danielsson, G. Ferretti and B. Sundborg, \IJMP {\bf
A11} (1996) 5463, hep-th/9603081.}
\lref\banks{T. Banks, W. Fischler, S. Shenker and L. Susskind, hep-th/9610043.}
\lref\horava{P. Horava and E. Witten, \NP {\bf B460} (1996) 506,
hep-th/9510209.}
\lref\dai{J. Dai, R.G. Leigh and J. Polchinski, \Mod {\bf A4} (1989)
2073.}
\lref\green{M. Green, J. Schwarz and E. Witten, ``Superstring
Theory,''
Cambridge University Press, 1987.}
\lref\witten{E. Witten, \NP {\bf B443} (1995) 85, hep-th/9503124.}
\lref\oldzero{M. Claudson and M.B. Halpern, \NP {\bf B250} (1985) 689;
M. Baake, P. Reinicke and V. Rittenberg, \JMP {\bf 26} (1985) 1070;
R. Flume, \AP {\bf 164} (1985) 189.}
\lref\doug{M.R. Douglas, D. Kabat, P. Pouliot and S.H. Shenker,
hep-th/9608024.}
\lref\bouwit{E. Witten, \NP {\bf B460} (1996) 335, hep-th/9510135.}
\lref\kabat{D. Kabat and P. Pouliot, \PRL {\bf 77} (1996) 1004,
hep-th/9603127.}
\lref\hoppe{B. de Wit, J. Hoppe, H. Nicolai, \NP {\bf B305} (1988)
545;
J. Hoppe, hep-th/9609232; J. Fr\"ohlich and J. Hoppe, hep-th/9701119.}
\lref\cecotti{S. Cecotti, P. Fendley, K. Intriligator and C. Vafa, \NP
{\bf B386} (1992) 405; S. Cecotti and C. Vafa, \CMP {\bf 158} (1993) 569.}
\lref\lowe{D.A. Lowe, to appear.}
\lref\dewit{B. de Wit, hep-th/9701169.}
\lref\nicolai{B. de Wit, M. L\"uscher and H. Nicolai, \NP {\bf B320}
(1989) 135.}

\newsec{Introduction}

Recently there has been much study of the strong coupling behavior
of ten-dimensional supersymmetric string theories, and their
simplest compactifications. A complex web of dualities has
emerged, unifying the ten-dimensional string theories and
relating them to the eleven-dimensional theory known as
M-theory. These strong/weak coupling dualities
predict correspondences between the perturbative states of one
string theory and nonperturbative bound states of another, allowing
stringent tests of the proposed dualities to be made.
In this paper we test Type I$'$/$E_8 \times E_8$ heterotic duality,
and Type I$'$/M-theory duality by studying the spectrum of
bound states of D(irichlet)-particles in the Type I$'$ theory.

The Type I' theory is T-dual to the usual Type I superstring theory
compactified to nine dimensions, and may be constructed
as an orientifold of Type IIA theory, which contains a background
of 16 D8-branes \dai. The Type I
theory in ten-dimensions is related via strong/weak coupling
duality to the $SO(32)$ heterotic string \refs{\witten, \polwit}.
The Type I$'$ theory is therefore dual to the $SO(32)$
heterotic theory compactified to nine dimensions and because the
two versions of the heterotic theory are related by T-duality below
ten dimensions \ginsparg, the Type I$'$ theory is related
via strong/weak coupling duality to the $E_8 \times E_8$ heterotic
theory.
This duality predicts a tower of D-particle bound states which
must match the perturbative spectrum of the
$E_8 \times E_8$ heterotic theory \kacsil\ (suitably broken by Wilson
lines).

The strongly coupled $E_8 \times E_8$ heterotic string has been
conjectured to be dual to a compactification of M-theory on
an interval $S^1/\BZ_2$ \horava. This is
related via a duality to a different
strong coupling limit of Type I$'$ theory. Again a tower
of D-particle bound states are predicted that may be checked
against the  M-theory spectrum.

The D-particle bound states in Type IIA string theory
are equivalent to supersymmetric ground states in the quantum
mechanics of a matrix model studied some time ago in \oldzero, and
more recently in \refs{\bouwit \kabat \danieltwo{--} \doug}. In these
works much was learned about the spectrum of non-BPS states,
and scattering of single-particle BPS states. The
existence of supersymmetric bound states of two or more D-particles has not
yet been established. In certain
truncations of the matrix quantum mechanics describing two
D-particles \hoppe, it has been
argued that a normalizable supersymmetric ground state does not
exist. Later we will comment on why the bound states considered in the
present paper do not survive the analog of the truncations used in
\hoppe, providing a possible explanation of these results.

D-particle bound states in Type I$'$ theory were originally considered
by Danielsson and Ferretti \daniel, where the matrix model describing
their dynamics
was constructed and non-BPS excitations and D-particle scattering
were studied. The appearance of the BPS bound states required by Type
I$'$/heterotic and Type I$'$/M-theory\foot{The M(atrix)-theory \banks\ limit of
the Type I$'$
D-particle matrix model is obtained by sending the D-particle number
to infinity. Following \banks\ this matrix model should provide
a definition of quantum eleven-dimensional M-theory on $S^1/\BZ_2$.}
duality was motivated in \kacsil.

In the present work, we continue the study of these bound states.
The direct construction of such states in the matrix model
formulation is in general a very difficult task. To make
progress, we will use a Born-Oppenheimer approximation. This
approach is similar to that of \daniel\ but differs in some
important details. The bound states that show up in this
approximation provide constraints on the quantum numbers
of the exact bound states. For low values of the D-particle
number these states are explicitly constructed, and the
quantum numbers found are consistent with
the proposed dualities.

\newsec{Predictions of Duality}

As discussed in \kacsil\ duality between the Type I$'$ theory and
$E_8\times E_8$ heterotic string theory compactified to
$9$ dimensions predicts the existence of certain bound states
of D-particles and D$8$-branes. This may be seen by compactifying the $SO(32)$
heterotic string theory on a circle with Wilson line
$(\half^8,0^8)$, in standard notation. This is related by $T$-duality
to a compactification
of the $E_8\times E_8$ heterotic string \ginsparg.  The
parameters of the two theories are related as
\eqn\hettyp{
\lambda_E = {R_{I'}^{3/2} \over \lambda_{I'}^{1/2}} ~,\qquad
R_E = \sqrt{R_{I'}\lambda_{I'}} ~.
}
The Wilson line breaks the gauge symmetry to $SO(16)\times SO(16)$.
In the limit that $R_E \to \infty$ with $\lambda_E$ fixed,
states which become massless
have even Kaluza-Klein momentum and give
the $N=1$ gravity multiplet with spacetime (i.e. $SO(8)$) quantum
numbers ${\bf 1+28+35_v+8_s+56_s}$ , plus the adjoint ${\bf
(1,120)+(120,1)}$ of
$SO(16)\times SO(16)$ in the ${\bf 8_v}$ and ${\bf 8_s}$;
or have odd Kaluza-Klein momentum and give spinor reps
${\bf (1,128)+(128,1)}$ in the ${\bf 8_v}$ and ${\bf 8_s}$.
 These multiplets combine to give $E_8\times E_8$
gauge bosons and the ten-dimensional $N=1$ supergravity multiplet
in the decompactification limit.
In terms of the Type I$'$ string, the states with Kaluza-Klein momentum $n$
map to bound states
of $n$ D-particles with the D8-branes. For $n$-even, we expect
to find the $N=1$ gravity multiplet, plus the ${\bf (1,120)+(120,1)}$
charged states; for $n$-odd we expect to find the  ${\bf (1,128)+(128,1)}$
states.

Both these theories may be viewed as different compactifications
of M-theory \horava\ on $S^1 \times S^1/\BZ_2$, with radii
$R_0$ and $R_1$. The relations between the parameters of the
theories are as follows
\eqn\metyp{
R_0= \lambda_{I'}^{2/3}= {R_E \over \lambda_E^{1/3} }~, \qquad
R_1={R_{I'}\over \lambda_{I'}^{1/3}} = \lambda_E^{2/3}~.
}

One may also consider the limit $R_0$ and $ R_1 \to \infty$, when
the massless fields should become the usual eleven-dimensional
supergravity
multiplet propagating in the bulk, together with massless $E_8\times
E_8$ gauge bosons propagating on the ends of the interval. In terms of
Type I$'$ parameters we have $R_{I'}\to \infty$ and $\lambda_{I'}\to
\infty$.
Duality between Type I$'$ and M-theory predicts bound states
of even numbers of D-particles in the
\eqn\gravm{
{\bf (8_v+8_s)\times(8_v+8_c)} = {\bf
1+28+35_v+8_v+56_v+8_s+8_c+56_s+56_c}~,
}
of $SO(8)$
will fill out the remainder of the
eleven-dimensional supergravity multiplet. The modes carrying zero
momentum in the $S^1/\BZ_2$ direction will be the same as in the
heterotic case, mentioned above. Bound states
with $SO(16)\times SO(16)$ charge
will give rise to $E_8\times E_8$ gauge multiplets, which only
propagate on the ends of the interval and lie in the ${\bf 8_v+8_s}$
as before.

\newsec{Bound States of D-particles and D8-branes}

Let us now consider these bound states directly in the Type I$'$
formalism. Here we will clarify a number of points not
addressed in \kacsil.
The Hamiltonian governing the $0$-$8$ brane system is discussed
in \daniel. This describes the situation where 8 D 8-branes are
sitting at each orientifold plane. This
system has $SO(16)\times SO(16)$ spacetime gauge symmetry, and
is dual to the heterotic string compactification discussed above.
Only in this limit is the Type I$'$
string coupling a constant as a function of the compactified
coordinate.

As discussed in \polwit\ , if we consider the situation
when $n$ 8-branes are at $x^9=0$ and $16-n$ are at $x^9=\pi R_{I'}$,
the Type I$'$ theory will develop a linear dilaton background,
provided the radius is sufficiently large. There exists
a critical radius at which the coupling of Type I$'$ diverges at
the end with the least number of 8-branes. In this limit
the D-particle bound states will become massless, and should make
up multiplets of $E_8\times E_8$. For the situation at hand,
the  dilaton is constant and $E_8\times E_8$ multiplets will  appear
in the infinite coupling limit for arbitrary values of the radius.

Let us review the $SO(N)$ symmetric matrix model \daniel\ governing
the low-energy dynamics of $N$ D-particles near one
of the orientifold planes. The bosonic worldline fields are as follows:
\eqn\bosons{
A_{0,9}^{IJ}~, \qquad X_i^{IJ}~, \qquad x_i^0~,
}
where $A$ is in the adjoint of $SO(N)$, $X$ is in the symmetric
traceless rep and $x_i^0$ is a singlet representing the position of
the
center of mass of the D-particles. Here $I,J=1,\cdots, N$ are
$SO(N)$ indices and $i=1,\cdots,8$ runs over the 8 noncompact spatial
dimensions. The fermionic modes are:
\eqn\fermions{
S_a^{IJ}~, \qquad S_{\dot a}^{IJ}~, \qquad s_{\dot a}~,
}
where $S_a$ is in the adjoint, $S_{\dot a}$ is in the traceless
symmetric rep and $s_{\dot a}$ is a singlet. Here $a$ and $\dot a$
label the spinor reps ${\bf 8_s}$ and ${\bf 8_c}$ of the eight
noncompact spatial dimensions. Finally there
are the degrees of freedom arising from open strings starting
at a D-particle and ending on one of the D$8$-branes, $\chi_r^I$ which
live in the fundamental of $SO(N)$. Here  $r=1,\cdots,16$ labels the
different 8-branes and their mirrors. The Hamiltonian
is given by
\eqn\hamil{
\eqalign{
H &= {\rm Tr}\biggl( \lambda_{I'}( \half P_i^2 - \half E_{9}^2)+
{1\over \lambda_{I'}}( \half [A_{9}, X_i]^2 - {1\over 4}[X_i, X_j]^2 )\cr &+
{i\over 2}(-S_a [|A_{9}|, S_a] - S_{\dot a} [|A_{9}|, S_{\dot a}] +
2 X_i \sigma^i_{a \dot a} \{S_a, S_{\dot a} \} ) \biggr) \cr &+
{i \over 2}(\chi_r^I | A_{9 IJ} |\chi_r^J +
\chi_r^I B_\mu^{rs} \chi_s^I )~,\cr}
}
in a gauge where $A_0$ has been set to zero. The $\sigma^i_{a \dot a}$
are defined as in \refs{\daniel, \danieltwo}.
The $B_\mu$ comes
from the coupling to the usual spacetime gauge bosons.
Note that fixing gauge $A_1=0$ requires us to impose the
Gauss law constraint on the space of physical states.

In fact,
this Hamiltonian is not quite correct. So far, we have ignored open string
winding modes which wrap the compactified direction (remember winding
number is conserved in Type I$'$ theory, while Kaluza-Klein momentum is
not).
The dynamics
of such modes may be ignored when the distance between the
orientifold planes is large. However, one effect of these
modes is to introduce a normal-ordering constant multiplying the terms
in the Hamiltonian linear in $A$. The first case in which
this term is nontrivial is the two D-particle case. In the following
we will compute this term and show it plays a crucial role in the
construction of the bound states.

For zero D-particle number, the massless states correspond to the
usual $N=1$ supergravity multiplet in nine dimensions together
with $SO(16)\times SO(16)$ gauge bosons arising from the usual $8-8$
open strings.

With just a single D-particle present
the Hamiltonian is trivial and independent of the fields $\chi$. The
quantum numbers of the states arise from quantizing the fermion
zero modes, as follows.
Define linear combinations of the $\chi_r$,
\eqn\newchi{
b_j = \half( \chi_j+i\chi_{j+8})~, \qquad b_j^*= \half(\chi_j - i
\chi_{j+8})~,
}
which satisfy the anticommutation relations
\eqn\bacom{
\{ b_i , b_j\}=0~,\quad \{ b_i^*, b_j^* \} =0~,\quad \{ b_i, b_j^* \} =
\delta_{ij}~.
}
The states $b^*_i |0\rangle$, $b^*_i b^*_j |0\rangle$ etc. give a
$2^8$ dim rep of $SO(16)$ which is the ${\bf 128+128'}$.
In addition, one must project out by an analog of the heterotic
GSO projection. In Type I$'$ this appears as a discrete $\BZ_2$ gauge
symmetry, which is a remnant of the gauge field on
the D-particle projected out by the orientifold projection.
On the $\chi$ fields, this will act as $(-1)^F$,
so to project onto gauge invariant states we keep only those
made up of even numbers of $\chi$'s -- yielding a single ${\bf 128}$ of
$SO(16)$. The other worldline fields are invariant under the discrete
gauge symmetry.
The spacetime quantum numbers arise by quantizing the
$s_{\dot a}$ zero modes. This is identical to the quantization of
fermion zero modes in the Green-Schwarz formulation of the Type I
string \green. The result is that a ${\bf 8_v}$ and ${\bf 8_s}$ of
$SO(8)$ appear.
This is in accord with Type I$'$/heterotic and Type I$'$/M-theory duality.

We now consider bound states of a pair of D-particles.
It is helpful to rewrite the general Hamiltonian \hamil\
for this particular case.
In terms of $SO(2)$ matrices, the $A_{9}$, $X_i$, $S_a$ and $S_{\dot a}$
fields are
\eqn\sotfie{
\eqalign{
A_{9}^{IJ} &= {i\over 2} \sigma_2 A\cr
X_i^{IJ} &= {1\over 2} (x_i \sigma_1 + \tilde x_i \sigma_3 ) \cr
S_a^{IJ} &= {i \over 2} \sigma_2 S_a \cr
S_{\dot a}^{IJ} &= {1\over 2} (S_{\dot a} \sigma_1+ \tilde S_{\dot a}
\sigma_3)~, \cr }
}
where the $\sigma^{IJ}_i$ are the usual Pauli matrices.
The $\chi$ fields live in the doublet of $SO(2)$, i.e.
$\chi=(\chi^1, \chi^2)$. Substituting
this into \hamil\ we find
\eqn\hamiltwo{
\eqalign{
H &= {1\over 4}\lambda_{I'} (p_i^2 + \tilde p_i^2+ p_A^2) +
{1\over \lambda_{I'}} \biggl(
{1\over 8} ( x_i \tilde x_j - \tilde x_i x_j)^2 +
 {1\over 4} A^2(x_i^2+\tilde x_i^2) \biggr) \cr  &
+{i\over 2} ( -\tilde S_{\dot a} |A |S_{\dot a} + \sigma^i_{a \dot a} S_a
(S_{\dot a} \tilde x_i  - \tilde S_{\dot a} x_i) )
+ {i \over 4} \chi^1 |A| \chi^2~.\cr}
}

The nontrivial (anti)commutation relations for the different fields are:
\eqn\commut{
\eqalign{
i[p_i, x_j] &= \delta_{ij}~, \qquad i[\tilde p_i, \tilde x_j] =
\delta_{ij}~,
\qquad i[p_A, A] = 1~, \qquad \{S_a , S_b \} = \delta_{ab}~, \cr
\{S_{\dot a}, S_{\dot b} \} &= \delta_{\dot a \dot b}~, \qquad
\{\tilde S_{\dot a}, \tilde S_{\dot b} \} = \delta_{\dot a \dot b}~, \qquad
\{ \chi^i_r, \chi^j_s \} = \delta^{ij} \delta_{rs}~. \cr}
}

To proceed further it is convenient to make a Born-Oppenheimer
approximation to study the two D-particle bound states. The
Hamiltonian only gives a correct description of D-particle
dynamics in the limit that their separation is smaller than
$l_s$, the string length. We may treat $A$ as a slowly fluctuating
mode and consider a limit
in which the length scale set by $A$ is less than $l_s$, but
much greater than the eleven-dimensional Planck scale $l_{pl}$.
This essentially corresponds to the moduli space approximation, and
was discussed in the Type II context in \doug. The approximation
breaks down when one considers ground states. The wavefunctions
of these states may depend sensitively on the $|A|< l_{pl}$ region.
However, because we know the wavefunction varies in a smooth
way as $|A|$ increases, the exact bound state wavefunctions should
match onto the approximate Born-Oppenheimer wavefunctions for
$|A|$ sufficiently large. This puts strong constraints on the
possible quantum numbers of the exact bound states, and we will find
it completely determines the $SO(16)$ gauge charges.

The Born-Oppenheimer approximation proceeds as follows.
The degrees of freedom are separated into the fast modes for
which the wavefunction is expected to vary rapidly, and the
slow modes for which the opposite is true.
For fixed values of the slow modes (which we take here to be
$A$, $S_a$ and the singlets $x_i^0$ and $s_{\dot a}$) we solve
for the eigenfunctions of the fast modes. In this limit the
$x_i$ and $\tilde x_i$ may be treated as ordinary harmonic
oscillators (the quartic piece in the $x$'s becomes
irrelevant). Likewise
the fermionic fields $S_{\dot a}$, $\tilde S_{\dot a}$ and the
$\chi_r$ should be treated as the Grassman analog of the harmonic
oscillator.

As mentioned above, there is an additional contribution to the
Hamiltonian coming from open string winding states. At low energies
these modes are in their ground states, but they may nevertheless
introduce extra normal-ordering terms into the Hamiltonian.
For us, the interesting modes are those which start on one
D-particle, wrap $n$ times around the $S^1$ and end on a
$8$-brane or the mirror D-particle. This
introduces an extra piece linear in $A$, which may be computed as follows.
First, let us define
\eqn\newchi{
\eqalign{
\chi_r &= {1\over \sqrt{2}} (\chi_r^1+  i\chi_r^2)~, \qquad
\bar \chi_r =
{1\over \sqrt{2}} (\chi_r^1- i \chi_r^2)~, \cr
\Sigma_{\dot a} &= {1\over \sqrt{2}} (S_{\dot a} + i \tilde S_{\dot a})~,
\qquad \bar \Sigma_{\dot a} = {1\over \sqrt{2}} (S_{\dot a} -
 i \tilde S_{\dot a} ) ~,\cr
\alpha_k &=  {1\over 2} \sqrt{A\over \lambda_{I'}}( x_k + i\tilde x_k)
+ {i \over 2}
\sqrt{\lambda_{I'}\over  A} (p_k+i \tilde p_k)~,
\cr
\bar \alpha_k &= {1\over 2} \sqrt{A\over \lambda_{I'}} (x_k -i\tilde x_k) +
{ i\over 2}
\sqrt{\lambda_{I'}\over A} (p_k-i\tilde p_k)~,\cr
}}
which allows us to write the terms in the Hamiltonian linear in $A$ as
\eqn\dham{
\delta H =\half |A|(\alpha_k^\dagger \alpha_k+ \bar
\alpha_k^\dagger\bar \alpha_k  +
\bar \Sigma_{\dot a} \Sigma_{\dot a}+\half \bar \chi_r \chi_r + c )~.
}
The determination of the constant term $c$ is reminiscent of the
determination of the normal-ordering constants in the physical state
conditions of heterotic string theory.
First we need to consider more carefully the coupling of
the winding modes to the gauge field $A$. The coupling is most
simply determined using the T-dual formulation of the D-string in
Type I \polwit. All the winding modes are periodic around the compact direction
except the winding modes of the $\chi$ fields, which are
antiperiodic.\foot{There is a different sector of the theory where the
winding modes of the
$\chi$ fields are periodic -- the analog of the NS sector of the
D-string.
This sector decouples from the low-energy physics.}
This implies the coupling of the winding modes to the $A$ field
takes the form
\eqn\hwind{
\delta H_{winding} = \sum_{n=1}^\infty \half |A|\biggl(
\alpha_{k,-n} \alpha_{k,
n}+
\bar
\alpha_{k,-n}\bar \alpha_{k,n}  +
n \Sigma_{\dot a,-n} \Sigma_{\dot a,n}+
(n-\half) \bar \chi_{r,-n+1/2} \chi_{r,n-1/2} \biggr)~,
}
where $n$ is the winding number and these modes obey the conventional
(anti)commutation relations \green. Thus we find that
$\delta H_{winding}$ is precisely proportional to
$L_0+\bar L_0$ of the Green-Schwarz formulation of the heterotic
string and
the normal-ordering constant is computed in the same way. Identifying
the $n=1$ modes with the fields in \dham\ in the obvious way, and
putting the rest of the winding modes in their ground states,
we recover \dham\ with $c=-1$.

The Gauss law constraint is likewise analogous to the
level-matching condition of the heterotic string --
the operator
\eqn\gauss{
C =\alpha_k^\dagger \alpha_k- \bar
\alpha_k^\dagger\bar \alpha_k  -
\bar \Sigma_{\dot a} \Sigma_{\dot a}+\half \bar \chi_r \chi_r -1~,
}
should annihilate physical states.

The lowest energy eigenstates of the Hamiltonian for the fast modes,
subject to this constraint are
\eqn\sstate{
| \psi \rangle = \bar \chi_r \bar \chi_s \prod_a S_a| 0\rangle~, \qquad
{\rm and} ~ \alpha_k^\dagger \prod_a S_a | 0\rangle~,
}
where $|0\rangle$ represents the ground state for all the bosonic
modes. Note
these states are even in the $\chi_r$ so are invariant under the
$\BZ_2$ discrete gauge symmetry.

To complete the Born-Oppenheimer approximation, we now quantize the
slow modes. To do this we identify $H_{slow} = \langle \psi |H | \psi
\rangle$. However this simply leaves us with $H_{slow} = \half \lambda_{I'}
p_A^2$. The lowest energy eigenfunction of this has wavefunction independent of
$A$. This state is normalizable since $A$ runs over the interval
$[0,2\pi R_{I'}]$. There are also other low-lying states with wavefunctions
\eqn\pslow{
\psi_{slow}(A) = {1\over \sqrt{\pi R_{I'}}} \cos({p A \over R_{I'}})~,
}
with $p$ an integer, corresponding to momentum in the $9$ direction.
Finally we must consider the fermion
zero modes $s_{\dot a}$ and $S_a$. The $s_{\dot a}$ behave as for
the single D-particle yielding spacetime quantum numbers ${\bf 8_v+
 8_s}$. In a similar way, the quantization of the $S_a$ modes
gives ${\bf 8_v+ 8_c}$. The final eigenstates are then a tensor
product of \sstate\ with the fermion zero mode states, yielding an
adjoint (${\bf 120}$) of $SO(16)$ with spacetime quantum numbers
\eqn\gmult{
(\bf 8_v+8_c)\times ( 8_v+  8_s)= {\bf 1 +  28+
35_v+8_v+56_v+8_s+8_c+56_s+56_c}~,
}
and a gauge singlet multiplet with spacetime quantum numbers
\eqn\gravmult{
\eqalign{
\bf 8_v \times (8_v+8_s) \times (8_v+8_c)&=  \bf
1+8_v+8_v'+8_v''+8_s+8_s'+8_c+8_c'+28+28'\cr & \bf +35_v+35_s+35_c +
56_v+56_s+56_s'+56_c+56_c'+112_v\cr &\bf +160_v+160_v'+160_s+160_c +224_{vs}+
224_{vc}+350~.\cr}
}

The states we have found include all the states predicted by Type
I$'$/heterotic duality and Type I'/M-theory duality.
We have also found additional bound states.
However we expect the degeneracy arising from the $S_a$ component
of the wavefunctions will be split when we go to next order in
the Born-Oppenheimer approximation due to the presence of the
$X S_a S_{\dot a}$ term in the Hamiltonian. The spacetime quantum
numbers of the exact bound states
may therefore be a subset of those appearing on the right-hand side of
\gmult\ and \gravmult. However, the gauge
charges are unaffected by this splitting since the $S_a$ are singlets
of $SO(16)$.

In an improved approximation, we
expect the states carrying gauge charge will have wavefunctions
peaked near $A=0$ with a width of order $l_{pl}$, and the states with
$p\neq 0$ will become massive. On the other hand, since gravitons
are free to propagate away from the orientifold plane, the whole
tower of gauge singlet states with $p\neq 0$ should become light in
the M-theory limit. Ideally, one would like to go beyond the Born-Oppenheimer
approximation and compute, for example, the Witten index,
which counts the number of
normalizable zero-energy states with $(-1)^F$ weighting. Work in
this direction is in progress \lowe.

We can see at this point how crucial it was that no truncations
of the supersymmetric quantum mechanics were made. The
zero-energy constraint and the Gauss law constraint take
a form very similar to the physical state conditions of the heterotic
string, which are sensitive to the spacetime dimension. This may
explain the results of \hoppe\ where no bound state was found in the
Type IIA context, when a truncation
to four spacetime dimensions was considered. Of course here it has
also been necessary to include an extra term in the Hamiltonian
arising from normal-ordering terms coupling to winding states--
such terms will not be present in the Type II version of this
calculation due to additional supersymmetry.

One could also have considered  different Born-Oppenheimer
approximations in which other combinations of the modes are treated
as slow. Apparently, the zero-point energy from the fast modes
in these other approximations is positive (barring any unexpected additional
terms in the Hamiltonian). This is not the case for the Type II
version of these calculations \refs{\hoppe,\nicolai,\dewit} where additional
supersymmetry leads to cancellation of the zero-point energies along
non-compact flat directions giving rise to a continuous spectrum, and
many associated subtleties.

It should be pointed out that the $0+1$-dimensional analog of the
jumping phenomena found in higher dimensional theories \cecotti\ may occur
as  the coupling $\lambda_{I'}$ and the radius $R_{I'}$ are varied.
In the preceding analysis we have assumed this does not occur, and
we found results consistent with duality. It would be interesting
to study this issue further to see whether the jumping conditions
on the heterotic side correspond to those on the Type I$'$ side.

\bigskip

\centerline{\bf Acknowledgments}

We wish to thank J. Park, J. Polchinski and J. Schwarz for helpful
discussions. This work
was supported in part by DOE grant DE-FG03-92-ER40701.

\listrefs
\end